\documentclass[12pt,preprint]{aastex}

\newcommand{\dqprot}{SDSS J1426}
\newcommand{\dqtwo}{SDSS J2348}
\newcommand{\dqone}{SDSS J2200}

\shorttitle{Two New Variable Hot DQ Stars}
\shortauthors{Barlow et al.}

\begin{document}

\title{Two New Variable Hot DQ Stars}

\author{B. N. Barlow$^{1,2}$\footnotetext[1]{Department of Physics and Astronomy, University of North Carolina, Chapel Hill, NC 27599-3255}\footnotetext[2]{Visiting astronomer, Cerro Tololo Inter-American Observatory, National Optical Astronomy Observatory, which is operated by the Association of Universities for Research in Astronomy, under contract with the National Science Foundation.}, B. H. Dunlap$^{1,2}$, R. Rosen\footnotetext[3]{National Radio Astronomy Observatory, P.O. Box 2, Green Bank, WV 24944}$^{3}$, \& J. C. Clemens$^{1,2}$ \\ \textit{Accepted by ApJ Letters on 8 October 2008}}

\begin{abstract}
We have discovered periodic variations in the light curves of two hot DQ stars from the Sloan Digital Sky Survey, SDSS J220029.08-074121.5 and SDSS J234843.30-094245.3.  These are the second and third variables detected among the hot DQs and confirm the existence of a new class of variable white dwarf stars.  The amplitudes of the variations are one half as large as those detected in the first discovered variable, SDSS J142625.71+575218.3, and required high signal-to-noise photometry to detect.  The pulse shapes of the stars are not like those of known white dwarf pulsators but are similar to the first hot DQ variable, SDSS J142625.71+575218.3.
\end{abstract}


\section{An Abundance of Variables Among the Hot DQs}
\label{intro}
Hot DQ stars are a recently-discovered class of white dwarfs with atmospheres composed primarily of carbon and devoid of significant amounts of hydrogen and helium \citep{duf07}.  Recently, \citet{mon08} announced the discovery of the first variable hot DQ star, SDSS J142625.71+575218.3 (hereafter \dqprot).  The star was one of six selected from the Sloan Digital Sky Survey (SDSS) and observed using the Argos high-speed photometer on the Otto Struve 2.1 m telescope at McDonald Observatory.  Based on theoretical arguments that couple the existence of non-radial oscillations with the presence of a surface convection zone in the models, \citet{mon08} identified \dqprot\ as the only likely pulsator.  In agreement with this expectation, the remaining five candidates did not exhibit variations at the 5 mma level.  

\citet{fon08} examined the asteroseismological potential of this class of white dwarfs using a full nonadiabatic approach.  They found that pulsational modes are excited in certain models with carbon-dominated atmospheres with the region of pulsational instability depending on the chemical composition, surface gravity, and effective temperature.  \citet{duf08a} used SDSS spectra and photometry to determine these parameters for a sample of nine hot DQ stars.  Based on these fits and using the general models of \citet{fon08}, \citet{duf08a} predicted that only \dqprot\ should exhibit pulsations.  To test this prediction, we initiated a program to observe the hot DQs not observed by \citet{mon08}.  Fortunately, we are in the process of commissioning the Goodman Spectrograph on the 4.1 m SOAR telescope and were able to apply several engineering nights to time-series photometric observations.  This observing technique is indispensable for diagnosing tracking, flexure, and image-quality problems in the telescope-instrument system.  Over the course of two runs in June and July 2008, we attempted to observe three hot DQ stars from the SDSS.  For one of these targets, we did not acquire a sufficient quantity of data under photometric conditions with a properly-functioning instrument and, therefore, cannot set useful limits on the existence of variations.  The other two stars show periodic variations with amplitudes around 7 mma, contrary to the expectations of \citet{duf08a}.

Our observations bring the total number of known variable hot DQs to three out of a sample of only eight studied.  This high fraction of variables and the failure of the predictions of \citet{duf08a} do nothing to strengthen the pulsational hypothesis.  The unusual harmonic structure and pulse shape of \dqprot\ suggested to \citet{mon08} that the star might be an interacting binary white dwarf akin to AM CVn \citep{pro95}, but with carbon-atmosphere white dwarfs.  In this case, the variations could be related to the binary orbital period or the precession of an accreting disk.  This model is attractive because it would neatly explain all three unusual properties of the star:  periodic variations, a carbon-dominated atmosphere, and broadened spectral lines.  It would also account for the high fraction of variables because the same mechanism that makes the atmosphere carbon-dominated requires some measure of variability.  Nevertheless, the presence of surface convection in the models at the right temperature is a compelling argument in favor of pulsations; every other known class of white dwarfs with surface convection zones exhibits non-radial pulsations.    

\section{Observations}
\label{obs}
The Goodman Spectrograph is an imaging spectrograph constructed for one of the Nasmyth ports of the SOAR 4.1 m telescope.  In imaging mode, which was the only mode employed for our observations, the camera-collimator combination re-images the SOAR focal plane with a focal reduction of approximately three times.  This yields a plate scale of 0.15 arcsec/pixel at the detector.  The camera contains a 4k x 4k Fairchild 486 back-illuminated CCD, with electronics and dewar provided by Spectral Instruments Inc. of Tucson, AZ.  The entire system is optimized for high throughput from 320 to 850 nm and uses optics of fused silica, CaF$_2$, and NaCl.  Details of the instrument can be found in \citet{cle04}.

We observed over two separate engineering weeks in June and July of 2008; a log of our observations is shown in Table 1.   Each data set consists of an uninterrupted run of time-series photometry, with the exception of the 27 July observation of SDSS J220029.08-074121.5 (hereafter \dqone), for which there is a five-minute gap in the data.  Although one data set for each object was obtained through an S8612 filter, the rest of our runs were unfiltered.  The S8612 filter has a bandpass of 300 to 700 nm, similar to BG40.  The CCD readout was unbinned for the 31 July observation of SDSS J234843.30-094245.3 (hereafter \dqtwo) but binned 2x2 for all other data.  In order to reduce the cycle time of the exposures, we restricted the readout of the CCD to a subsection of the detector and used an intermediate readout speed (100 kHz).  The seeing averages for the runs ranged from 1" to 2.5", and there was significant moonlight in many of the runs.

For the runs acquired in June, there were three problems with the spectrograph.  The first of these was a periodic pattern in the bias frames with an amplitude of approximately 20 electrons.  The second was poor control over shutter timing, rendering the FITS header time stamps accurate only to one second from UTC as measured by the facility GPS.  Finally, the CCD subsection used in the \dqone\ frames contained a rectangular area of higher counts spanning the length of the serial direction and around 100 pixels in the parallel direction.  In all cases, we positioned the target well outside of this region and were cautious not to employ comparison stars near the boundary of this anomalous section of the chip.  The primary objective of our engineering was to resolve these problems, and the discoveries presented in this Letter are a byproduct of this effort.  Before the July run, we diagnosed and eliminated the bias noise, but the the other two issues are still under study.  We do not believe any of these problems produced misleading artifacts in the data; nonetheless, we have taken care to present for each variable candidate the amplitude spectrum of a comparison star reduced in the same way as the candidate photometry.  In no case do the comparison stars show significant periodic signals.  Moreover, the variable candidates showed the same oscillation frequencies in all of the observations.

\section{Reduction and Analysis}
\label{analysis}
\begin{figure}
\epsscale{1}
\plotone{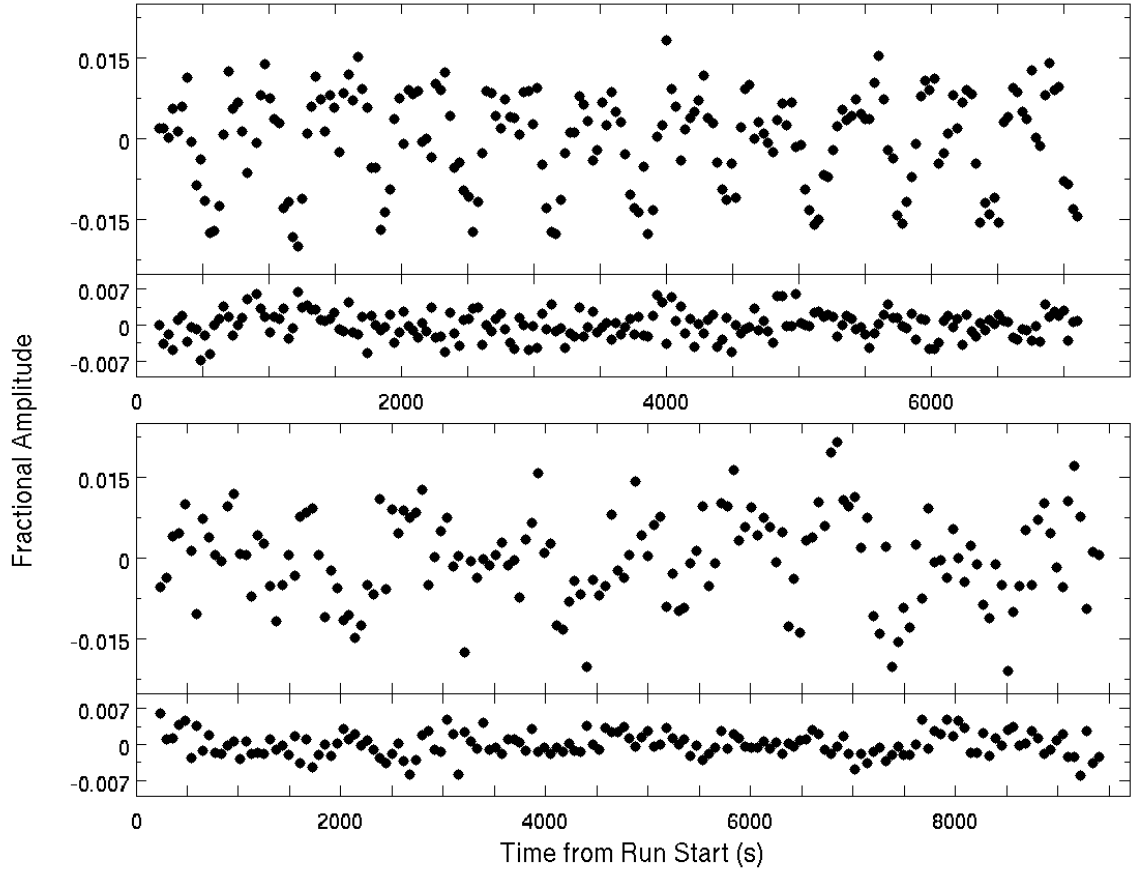}
\caption{Representative light curves for \dqone\ (top panel) and \dqtwo\ (bottom panel) shown for the nights of 28 Jun and 31 Jul, respectively.  The data are presented unsmoothed.}
\label{fig:lc}
\end{figure}

We extracted our photometry using the external IRAF package CCD\_HSP developed by Antonio Kanaan, which employs the aperture photometry preferred by \citet{odo00}.  The extraction parameters used maximized the S/N ratio in the light curves.  The aperture widths ranged from 1.7 to 2 times the seeing width, while sky annuli started approximately 3" from the center of the stars and had widths around 1.5".  To correct for small-scale transparency variations, we divided our light curves by those of constant comparison stars and normalized the resulting curves with parabolas.  We did not flat-field or bias-subtract any of the frames, due to superb telescope guiding and the instrumental issues mentioned in \S\ref{obs}.  Unsmoothed light curves of the longest runs for \dqtwo\ and \dqone\ are presented in Figure \ref{fig:lc}. 

\begin{figure}
\epsscale{1}
\plotone{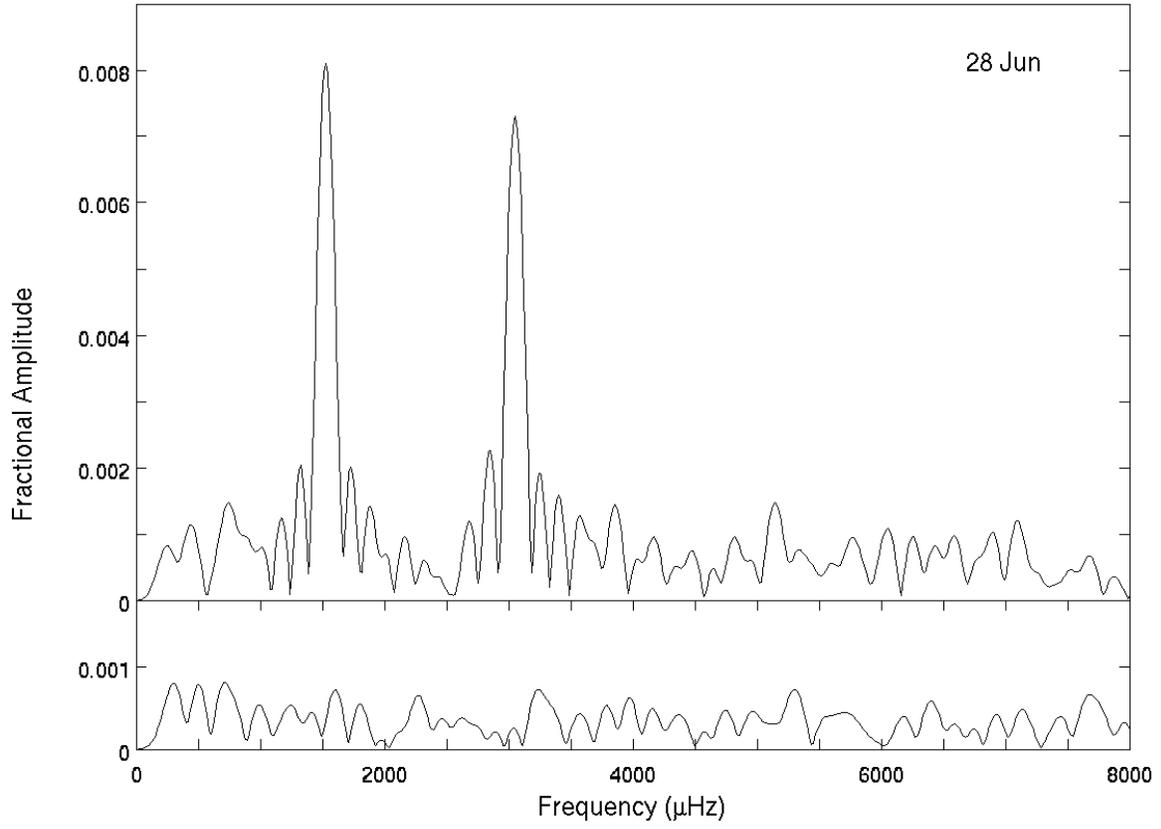}
\caption{Representative amplitude spectrum of \dqone\ (top) and a nearby comparison star (bottom).}
\label{fig:dft1}
\end{figure}

\begin{figure}
\epsscale{1}
\plotone{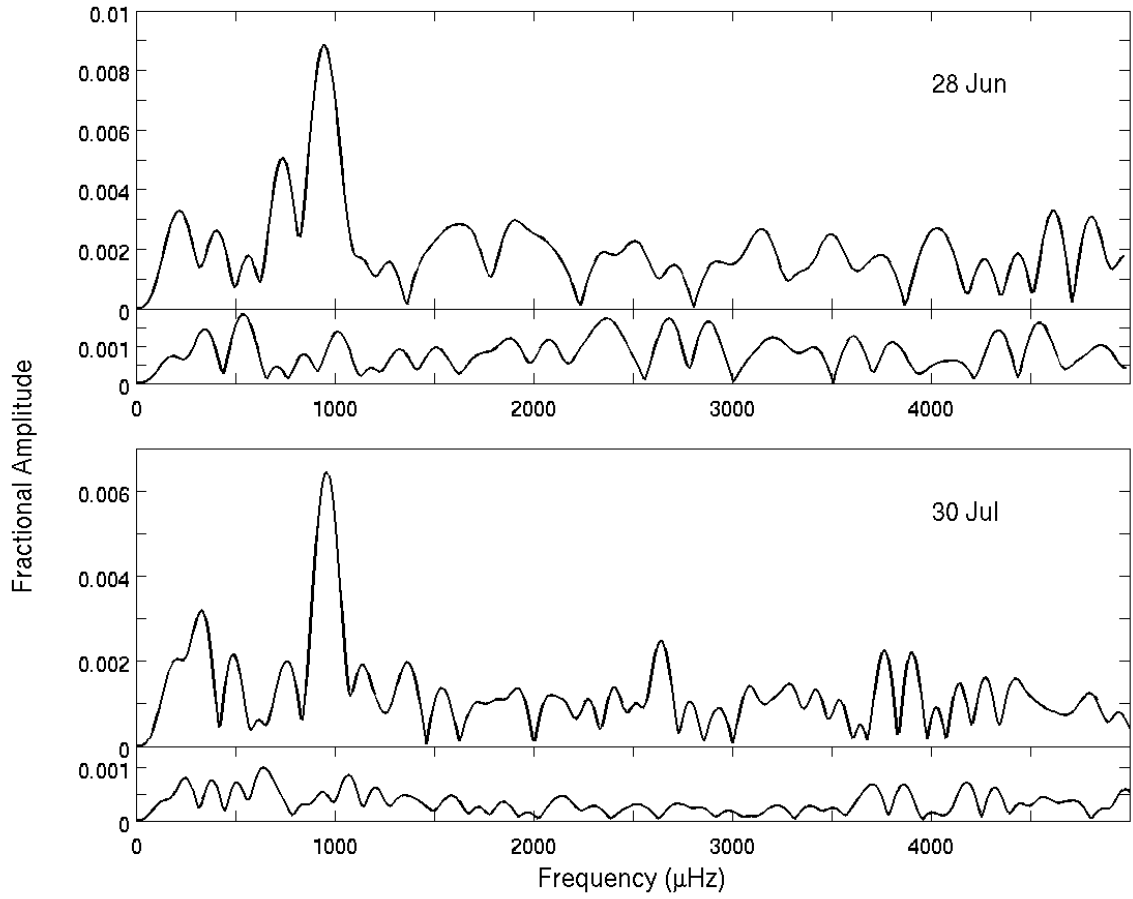}
\caption{Amplitude spectra of \dqtwo\ (top half of each panel) and a nearby comparison star (bottom half of each panel).}
\label{fig:dft2}
\end{figure}

We analyzed the reduced light curves using two tools:  the discrete Fourier transform and least-squares fitting of sine waves to the light curves$^{4}$\footnotetext[4]{These tools were employed via the WQED suite (http://www.physics.udel.edu/darc/wqed/index.html)}.  Figures \ref{fig:dft1} and \ref{fig:dft2} present amplitude spectra for \dqone\ and \dqtwo, respectively, displayed above those for nearby comparison stars reduced in the same manner.  Table 2 summarizes the parameters derived from the least-squares fittings.  The errors shown in Table 2 are from the least-squares fittings and implicitly assume uncorrelated noise.  For correlated noise, as produced by transparency variations in the atmosphere, the errors may be three times as large (see \citealt{mon99}).  

The amplitude spectrum of \dqone\ exhibits a large signal near 1524 $\mu$Hz (656 s) and its first harmonic (3055 $\mu$Hz).  The probability that a signal this large could occur by chance (the ``false alarm'' probability, \citealt{hor86}),  is less than 10$^{-14}$.  Moreover, a time span of more than four weeks separates the first and last observations, and the frequency measurements agree to within the errors.  The amplitude spectra do not show any other statistically significant periodicities, nor are there any in the comparison stars.

\dqtwo\ is 1.3 \textit{g} magnitudes fainter than \dqone, so the noise in the amplitude spectra is larger.  Nonetheless, we detect a variation near 951 $\mu$Hz (1052 s) in both runs (see Figure 2).  In addition, there is a large sidelobe in the 28 June run and a formally significant signal at 2641 $\mu$Hz (379 s) in the 31 July run.  The sidelobe corresponds in frequency to a feature in the window function (not shown), so we cannot regard it as the detection of an independent frequency.  The signal at 379 s is not harmonically related to the 1052 s variation or any window feature and is potentially suggestive of multi-periodic pulsations.  The false alarm probability for this frequency is 15\% (as a comparison, the largest peak has a false alarm probability less than 10$^{-6}$).  However, this signal appears in only one of our two runs and does not behave like a variation in the star should when we degrade the signal-to-noise ratio.  We do this by successively increasing the aperture size used in the photometry and recalculating the amplitude spectrum.  The amplitude of the 1052 s variation remains approximately constant as the noise around it increases, but the amplitude of the 379 s signal increases significantly with larger aperture size.  Therefore, we cannot claim to have detected a second independent frequency in the light curve of this star, however interesting such a detection might be.

The large harmonic present in \dqone\ suggests it would be fruitful to look at the folded pulse shape. In Figure 4 we show folded light curves for both objects.  The pulse shape for \dqone\ resembles the pulse shape for \dqprot\ (see Figure 4 of \citealt{mon08}) more than it resembles that of a typical pulsating white dwarf.  The shape for \dqtwo\ looks like the light curve of a ZZ Ceti when we plot it upside-down (see Figure 5).  This reinforces the point made by \citet{mon08} that while in a typical white dwarf pulsator the harmonic phase makes the peaks higher and the valleys shallower, in the hot DQ variables the phase makes the peaks lower and valleys deeper.  While this does not rule out pulsation, it does add the requirement that any pulsational model explain this difference. 

\section{Two Proposed Mechanisms Remain Viable}
\label{dis}

We have conducted high signal-to-noise time-series photometry for two hot DQs not studied by \citet{mon08}.  Two out of two targets observed with a 4 m telescope show periodic variations.  Overall, three out of eight targets studied to date show variations.  This raises a question about this class of stars: does that which makes them hot DQs also make them vary?  If they are interacting binaries transferring C or C-O, then the unusual composition and the photometric variations are caused by the same phenomena.  If they are pulsators, \citet{duf08c} have also proposed a connection between the surface composition and the variations through the existence of convection in the models.  They propose that the C convection zone that develops mixes and dilutes a thin, overlying He layer.  The same convection could drive pulsations \citep{bri90,gol99}.  A problem with this picture is raised by \citet{duf08b}, who report the presence of a strong magnetic field in \dqprot.  They note that a magnetic field could supress convection and, perhaps, pulsation, and this has not been treated self-consistently in the pulsation models.  In either case, the results of this Letter show that the pulsation models have not yet yielded a predictive success.  Whether this is a consequence of poor atmospheric parameter determinations or whether it means the variable hot DQs are not pulsators is unclear.  At this moment, there is no \textit{a priori} reason to prefer pulsations over other explanations for hot DQ variations.  The best evidence for pulsations would be the convincing detection of multiple incommensurate frequencies, and we are pursuing observations to further increase the signal-to-noise ratio in the amplitude spectra.  Citing as yet unpublished data (E.M. Green 2008, in preparation), \citet{duf08b} report a probable detection of an incommensurate period in \dqprot, which could strengthen the case for pulsations.

As an alternative to pulsations, the interacting binary white dwarf hypothesis proposed by \citet{mon08} has lost none of its appeal.  \citet{mon08} have already pointed out that the pulse shape in \dqprot\ resembles AM CVn, a known interacting binary white dwarf.  Likewise, the pulse shapes of \dqone\ in our Figure 4 are reminiscent of V803 Cen \citep{odo89}, a binary whose amplitude spectrum shows a fundamental at 1611 s and multiple harmonics.  Interestingly, it also shows an incommensurate period at 175 s whose origin is unclear.  Clearly, time-series photometry alone will not be sufficient to arrive at a definitive model for the hot DQ variables.  The binary hypothesis also offers an attraction the pulsation models do not; it predicts that a large fraction of the hot DQs will be variables, which is what we have found in our sample.  Conversely, it does not directly predict that the stars will have similar temperatures, and so the pulsational interpretation is more successful in explaining the temperature clustering (see \citealt{duf08c}) of the hot DQ stars.

\acknowledgements
We acknowledge the support of the National Science Foundation, under award AST-0707381, and are grateful to the Abraham Goodman family for providing the financial support that made the spectrograph possible.  We thank the Delaware Asteroseismic Research Center for providing the filter used in these studies.  We also recognize the observational support provided by the SOAR operators Alberto Pasten, Patricio Ugarte, Sergio Pizarro, and Daniel Maturana.  The SOAR Telescope is operated by the Association of Universities for Research in Astronomy, Inc., under a cooperative agreement between the CNPq-Brazil, the National Observatory for Optical Astronomy (NOAO), the University of North Carolina, and Michigan State University.  Finally, we thank an anonymous referee for helping us clarify our presentation in various parts of this Letter.

\clearpage

\begin{table}
\caption{Observations Log}
\begin{tabular}{ccccccc}
\hline
\hline
Object & UT Date & Start Time & T$_{exp}$ & T$_{cycle}$ & Length & Filter\\
 & (2008) & (UTC) & (s) & (s) & (s) & \\
\hline
\dqone & 27 Jun & 06:57:56 & 30 & 34.8 & 7314 & none\\
 & 28 Jun & 06:00:00 & 30 & 34.8 & 7380 & S8612\\
 & 27 Jul & 02:31:44 & 25 & 29.6 & 6170 & none\\
\dqtwo & 28 Jun & 08:47:18 & 90 & 95.5 & 7446 & S8612\\
 & 31 Jul & 05:57:30 & 55 & 59.5 & 9520 & none\\
\hline
\end{tabular}
\label{table:obs}
\end{table}

\clearpage

\begin{table}
\caption{Periodicites detected in the light curves}
\begin{tabular}{ccccc}
\hline
\hline
Object & UT Date (2008) & Frequency ($\mu$Hz) & Amplitude (mma) & Phase\tablenotemark{a} (cycles)\\
\hline
\dqone & 27 Jun & 1531.2 $\pm$ 8.7 & 6.68 $\pm$ 0.75 & 0.88 $\pm$ 0.02\\
 &  & 3053.3 $\pm$  7.1 & 7.60 $\pm$ 0.71 & 0.48 $\pm$ 0.02\\
 & 28 Jun & 1527.4 $\pm$ 6.5 & 8.04 $\pm$ 0.66 & 0.09 $\pm$ 0.01\\
 & & 3052.9 $\pm$ 7.6 & 7.30 $\pm$ 0.70 & 0.89 $\pm$ 0.02\\
 & 27 Jul & 1512 $\pm$ 11 & 7.40 $\pm$ 0.93 & 0.29 $\pm$ 0.02\\
 & & 3060 $\pm$ 11 & 7.26 $\pm$ 0.94 & 0.41 $\pm$ 0.02\\
\dqtwo & 28 Jun & 944 $\pm$ 12 & 8.8 $\pm$ 1.3 & 0.27 $\pm$ 0.02 \\
 & 31 Jul & 958 $\pm$ 6.7 & 6.4 $\pm$ 0.8 & 0.38 $\pm$ 0.02 \\
\hline
\tablenotetext{a}{The uncertainties shown are lower limits.}
\end{tabular}
\label{table:freqsol}
\end{table}

\clearpage

\clearpage

\clearpage

\clearpage

\begin{figure}
\epsscale{1}
\plotone{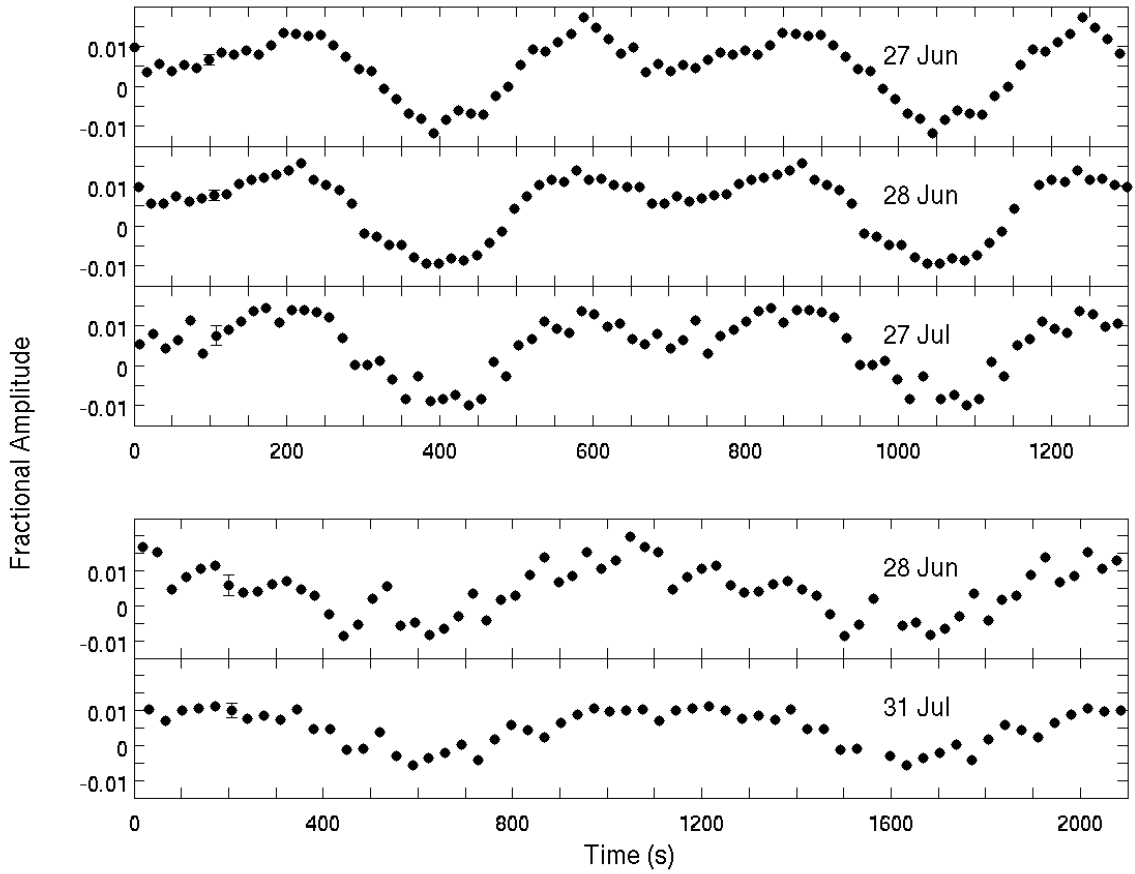}
\caption{Folded pulse shapes for \dqone\ (top panel) and \dqtwo\ (bottom panel).  Representative error bars are shown for points near 100 s (top panel) and 200 s (bottom panel).}
\label{fig:flc}
\end{figure}

\clearpage

\begin{figure}
\epsscale{1}
\plotone{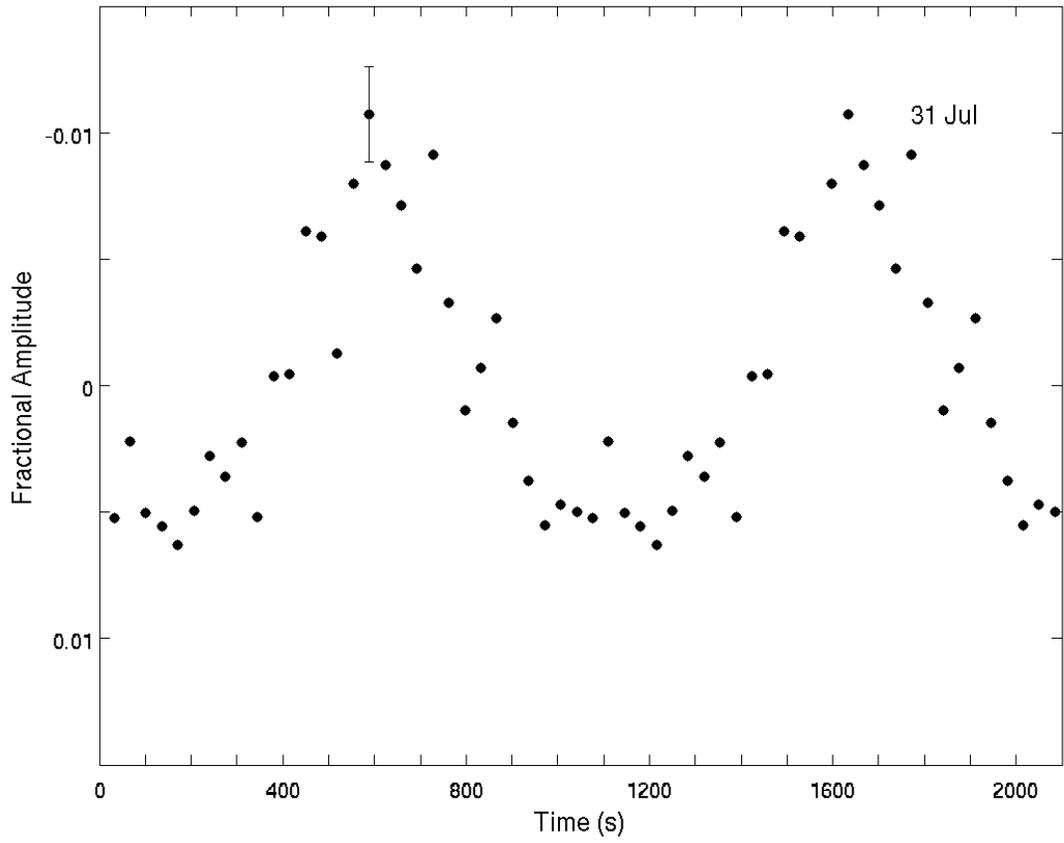}
\caption{Upside-down folded pulse shape for \dqtwo\ from 31 July.  A representative error bar is shown for the point near 600 s.}
\end{figure}



\begin{thebibliography}{14}
\expandafter\ifx\csname natexlab\endcsname\relax\def\natexlab#1{#1}\fi

\bibitem[{{Brickhill}(1990)}]{bri90}
{Brickhill}, A.~J. 1990, \mnras, 246, 510

\bibitem[{{Clemens} {et~al.}(2004){Clemens}, {Crain}, \& {Anderson}}]{cle04}
{Clemens}, J.~C., {Crain}, J.~A., \& {Anderson}, R. 2004, in Presented at the
  Society of Photo-Optical Instrumentation Engineers (SPIE) Conference, Vol.
  5492, Ground-based Instrumentation for Astronomy. Edited by Alan F. M.
  Moorwood and Iye Masanori. Proceedings of the SPIE, Volume 5492, pp. 331-340
  (2004)., ed. A.~F.~M. {Moorwood} \& M.~{Iye}, 331--340

\bibitem[{{Dufour} {et~al.}(2008{\natexlab{a}}){Dufour}, {Fontaine}, {Liebert},
  {Schmidt}, \& {Behara}}]{duf08a}
{Dufour}, P., {Fontaine}, G., {Liebert}, J., {Schmidt}, G.~D., \& {Behara}, N.
  2008{\natexlab{a}}, ArXiv e-prints, 805

\bibitem[{{Dufour} {et~al.}(2008{\natexlab{b}}){Dufour}, {Fontaine}, {Liebert},
  {Williams}, \& {Lai}}]{duf08b}
{Dufour}, P., {Fontaine}, G., {Liebert}, J., {Williams}, K., \& {Lai}, D.~K.
  2008{\natexlab{b}}, ArXiv e-prints, 807

\bibitem[{{Dufour} {et~al.}(2007){Dufour}, {Liebert}, {Fontaine}, \&
  {Behara}}]{duf07}
{Dufour}, P., {Liebert}, J., {Fontaine}, G., \& {Behara}, N. 2007, \nat, 450,
  522

\bibitem[{{Dufour} {et~al.}(2008{\natexlab{c}}){Dufour}, {Liebert}, {Fontaine},
  \& {Behara}}]{duf08c}
{Dufour}, P., {Liebert}, J., {Fontaine}, G., \& {Behara}, N.
  2008{\natexlab{c}}, in Astronomical Society of the Pacific Conference Series,
  Vol. 391, Hydrogen-Deficient Stars, ed. A.~{Werner} \& T.~{Rauch}, 241--+

\bibitem[{{Fontaine} {et~al.}(2008){Fontaine}, {Brassard}, \& {Dufour}}]{fon08}
{Fontaine}, G., {Brassard}, P., \& {Dufour}, P. 2008, \aap, 483, L1

\bibitem[{{Goldreich} \& {Wu}(1999)}]{gol99}
{Goldreich}, P., \& {Wu}, Y. 1999, \apj, 511, 904

\bibitem[{{Horne} \& {Baliunas}(1986)}]{hor86}
{Horne}, J.~H., \& {Baliunas}, S.~L. 1986, \apj, 302, 757

\bibitem[{{Montgomery} \& {O'Donoghue}(1999)}]{mon99}
{Montgomery}, M.~H., \& {O'Donoghue}, D. 1999, Delta Scuti Star Newsletter, 13,
  28

\bibitem[{{Montgomery} {et~al.}(2008){Montgomery}, {Williams}, {Winget},
  {Dufour}, {DeGennaro}, \& {Liebert}}]{mon08}
{Montgomery}, M.~H., {Williams}, K.~A., {Winget}, D.~E., {Dufour}, P.,
  {DeGennaro}, S., \& {Liebert}, J. 2008, \apjl, 678, L51

\bibitem[{{O'Donoghue} {et~al.}(2000){O'Donoghue}, {Kanaan}, {Kleinman},
  {Krzesinski}, \& {Pritchet}}]{odo00}
{O'Donoghue}, D., {Kanaan}, A., {Kleinman}, S.~J., {Krzesinski}, J., \&
  {Pritchet}, C. 2000, Baltic Astronomy, 9, 375

\bibitem[{{O'Donoghue} \& {Kilkenny}(1989)}]{odo89}
{O'Donoghue}, D., \& {Kilkenny}, D. 1989, \mnras, 236, 319

\bibitem[{{Provencal} {et~al.}(1995){Provencal}, {Winget}, {Nather},
  {Robinson}, {Solheim}, {Clemens}, {Bradley}, {Kleinman}, {Kanaan}, {Claver},
  {Hansen}, {Marar}, {Seetha}, {Ashoka}, {Leibowitz}, {Meistas}, {Bruvold},
  {Vauclair}, {Dolez}, {Chevreton}, {Barstow}, {Sansom}, {Tweedy}, {Fontaine},
  {Bergeron}, {Kepler}, {Wood}, \& {Grauer}}]{pro95}
{Provencal}, J.~L., {Winget}, D.~E., {Nather}, R.~E., {Robinson}, E.~L.,
  {Solheim}, J.-E., {Clemens}, J.~C., {Bradley}, P.~A., {Kleinman}, S.~J.,
  {Kanaan}, A., {Claver}, C.~F., {Hansen}, C.~J., {Marar}, T.~M.~K., {Seetha},
  S., {Ashoka}, B.~N., {Leibowitz}, E.~M., {Meistas}, E.~G., {Bruvold}, A.,
  {Vauclair}, G., {Dolez}, N., {Chevreton}, M., {Barstow}, M.~A., {Sansom},
  A.~E., {Tweedy}, R.~W., {Fontaine}, G., {Bergeron}, P., {Kepler}, S.~O.,
  {Wood}, M.~A., \& {Grauer}, A.~D. 1995, \apj, 445, 927

\end{thebibliography}
\end{document}